\documentclass[aps,prc,superscriptaddress,reprint]{revtex4-1}
\usepackage{tabularx}
\usepackage{graphicx} 
\usepackage[strings]{underscore}
\usepackage{CJKutf8}
\usepackage{epstopdf}
\usepackage{dcolumn} 
\usepackage{bm}
\usepackage{hyperref}[colorlinks, linkcolor=blue, anchorcolor=blue, citecolor=blue]
\usepackage{array}
\usepackage{setspace}
\usepackage{booktabs}
\usepackage{amsmath}
\usepackage{mathtools}
\usepackage[section]{placeins}
\hypersetup{
    colorlinks=true,
    linkcolor=blue,
    filecolor=gray,
    urlcolor=blue,
    citecolor=blue,
}
\begin{document}
\begin{CJK}{UTF8}{gbsn}

\title{Synthesis mechanism of superheavy element 120: a dinuclear system model approach with microscopic inputs}

\author{Wei Zhang(张炜)}
\affiliation{School of Physics, Zhengzhou University, Zhengzhou 450001, China}

\author{Shi-Jie Zhang(张士杰)}
\affiliation{School of Physics, Zhengzhou University, Zhengzhou 450001, China}

\author{Peng-Hui Chen(陈鹏辉)}
\email{Corresponding author: chenpenghui@yzu.edu.cn}
\affiliation{School of Physical Science and Technology, Yangzhou University, Yangzhou 225002, China}

\date{\today}
\begin{abstract}
	The dinuclear system model incorporates several essential input physical quantities, including nuclear mass, fission barrier, shell correction energy, level density parameter, and shell damping factor, etc., which are derived from diverse nuclear structure models. To achieve theoretical consistency, we try to generate these essential input physical quantities from the finite-temperature covariant density functional theory using PC-PK1 energy density functional, with pairing correlations treated via the BCS approach. With microscopically determined input parameters, the dinuclear system model can successfully reproduce experimental results for:  
	(i) cold fusion reaction systems ($^{48}$Ca + $^{204,206-208}$Pb $\rightarrow$ $^{252,254-256}$No$^*$), and 
	(ii) hot fusion reaction systems ($^{48}$Ca + $^{239,240,242,244}$Pu $\rightarrow$ $^{287,288,290,292}$Fl$^*$). Furthermore, we perform calculations for the fusion reactions $^{50}$Ti+$^{249}$Cf, $^{51}$V+$^{249}$Bk, $^{54}$Cr+$^{248}$Cm, and $^{55}$Mn+$^{243}$Am, targeting the synthesis of element 120. It is  found that the maximum synthesis cross section for these four reactions are 48.20 fb, 12.33 fb, 5.25 fb, 0.47 fb corresponding to $^{50}$Ti($^{249}$Cf,4n)$^{295}$120 at $E^*_{\rm CN}$ = 41 MeV, $^{51}$V($^{249}$Bk,3n)$^{297}$120 at $E^*_{\rm CN}$ = 34 MeV, $^{54}$Cr($^{248}$Cm,3n)$^{299}$120 at $E^*_{\rm CN}$ = 32 MeV, $^{55}$Mn($^{243}$Am,5n)$^{293}$120 at $E^*_{\rm CN}$ = 53 MeV, respectively.

\end{abstract}

\maketitle

\section{Introduction}

The synthesis of new superheavy elements has attracted a great deal of attention from theorists and experimentalists in the nuclear physics field, since the “island of stability” of superheavy nuclei (SHN) was introduced by the shell model in the 1960s \cite{Caurier:2004gf,physics-54-9-599}. Synthesizing these new SHN is critical to expanding our nuclide map and reaching these elusive islands of stability \cite{Tarasov:2024tsf,WOS:001182586900017}, which promise to reveal a deeper understanding of the fundamental nature of nuclear matter.

Laboratories around the world, such as the Institute of Modern Physics (IMP) of the Chinese Academy of Sciences (CAS) \cite{Gan:2022pox}, Gesellschaft f\"{u}r Schwerionenforschung (GSI) \cite{Heinz:2022aoq,Block:2022rud}, the Joint Institute for Nuclear Research (JINR) \cite{Oganessian2022}, the Rikagaku Kenky$\overline{\rm u}$sho (RIKEN) \cite{WOS:000453261100004,Tanaka:2020abd}, and the Lawrence Berkeley National Laboratory (LBNL) \cite{Pore:2020qqd}, have invested heavily in building or updating heavy ion accelerator and related detection equipment to explore the superheavy island of stability. Over the past half-century, experimentalists have synthesized and identified 15 superheavy elements with atomic numbers $Z$ ranging from 104 to 118 \cite{Thoennessen:2012mk}. Thus the current competition in synthesizing new SHN with atomic numbers $Z$=119 and 120 is very fierce. However, due to the extremely small cross sections for the synthesis of SHN, their experimental production remains a challenging task that is both time-consuming and costly. Therefore, reliable theoretical studies guiding experiments are crucial.

A number of theoretical models have been developed to study the synthesis of SHN, each providing a unique perspective on fundamental physics. These theoretical models include the time-dependent Hartree-Fock (TDHF) approach \cite{Maruhn:2013mpa,Guo:2018rkm,Sekizawa:2019gza}, the improved quantum molecular dynamics (ImQMD) model \cite{Wang:2002ywa,Jiang:2013jsa,Zhao:2013ena}, the GRAZING model based on dynamic analysis \cite{Yanez:2015boa}, the dinuclear system (DNS) model \cite{Feng:2006zeg,Feng:2007ycq,Zhu:2014jaa,Bao:2015gua,Adamian:2020zuo}, Langevin-type dynamical equations \cite{Shen2011,Boilley2011}, fusion-by-diffusion model \cite{Cap2011,Cap2022}, the two-step model \cite{Shen2002}, and several empirical approaches \cite{Loveland2007,Loveland2015}, etc. 
The DNS model has been enhanced to incorporate shell effects, dynamical deformation, fission, quasi-fission, deep-inelastic collisions, and the odd-even effect \cite{Feng:2006zeg,Bao:2015gua,Feng:2007ycq,FENG201082,PhysRevC.85.041601,PhysRevC.89.037601,PhysRevC.107.014616}. As a result, the model demonstrates a strong capability to accurately reproduce available experimental data and possesses nice predictive power\cite{2023chen}. However, the predictive accuracy of the model and the uncertainties in its outcomes are critically dependent on the underlying physical inputs \cite{PhysRevC.109.054611,PhysRevC.109.044618}. Moreover, essential input physical quantities—such as masses, fission barriers, and shell correction energies—used in the DNS model to calculate synthesis cross sections are derived from distinct theoretical nuclear models, potentially introducing redundancy in the treatment of certain physical factors. In previous calculations, nuclear masses were typically taken from the finite-range droplet model (FRDM) \cite{MOLLER20161}, fission barriers were derived from macroscopic-microscopic model \cite{MYERS1974186}, and shell correction energies were computed using the Strutinsky method based on single-particle levels from a Folded-Yukawa potential \cite{STRUTINSKY1967420,STRUTINSKY19681}. The level density parameters were usually assigned an empirical value, taken as \( a = A/12 \) \cite{PhysRevC.62.064303}.
Among many theoretical models describing the structure of atomic nuclei, covariant density functional theory (CDFT) has attracted much attention for the advantages of its inherent Lorentz covariance and has achieved great success in describing many properties of spherical and deformed nuclei all over the nuclear chart \cite{Ring:1996qi,Zhou:2003iu,Vretenar:2005zz,Meng:2005jv,Meng:2006wt,Liang:2014dma}.
For example, the pseudospin symmetry, the charge symmetry breaking, and nuclear level density are well described by CDFT recently \cite{Sun2023,Sun2024,Sun2025,Z2023,J2024}.
The dynamical extension of the CDFT, namely the time-dependent CDFT (TD-CDFT), has been developed to describe reaction dynamics \cite{Ren2020,Zhang024316,Zhang024614,Zhang2025}.
As a microscopic model, the CDFT with the BCS approach is employed to provide the essential physical inputs for the DNS model, going towards a self-consistent structure-based reaction framework.

In this study, to test the compatibility of microscopic input physical quantities with the DNS model, the synthesis cross sections for No and Fl isotopes are calculated and compared with experimental data. Based on this, we predict the synthesis cross sections of element 120. In detail, the single neutron separation energies $S_{\mathrm{n}}$, fission barriers $B_{\mathrm{f}}(E^*)$, shell correction energies $E_{\mathrm{sh}}$, as well as the asymptotic level density parameters $\tilde{a}$ of the compound nuclei $^{248-256}$No, $^{278-292}$Fl, and $^{290-302}$120 extracted from the finite-temperature CDFT are used as inputs to the DNS model. 

The structure of this article is as follows: Section \ref{sec2} provides a concise overview of the finite-temperature CDFT and DNS model. Section \ref{sec3} presents the computational findings and ensuing discussions. Finally, Section \ref{sec4} concludes with a summary of the key points.

\section{Theoretical framework}\label{sec2}

In this framework, the essential inputs for the DNS model to compute the synthesis cross-sections of SHN in fusion-evaporation reactions is provided based on the finite-temperature CDFT.

\subsection{Finite-temperature CDFT}

The starting point of the finite-temperature CDFT is an effective Lagrangian density
with zero-range point-coupling interaction between nucleons:
\begin{eqnarray}\label{Eq:Lagrangian}
	\mathcal{L}&=& \bar\psi(i\gamma_\mu\partial^\mu-m)\psi 
	-\frac{1}{2}\alpha_S(\bar\psi\psi)(\bar\psi\psi) \nonumber\\
	&& -\frac{1}{2}\alpha_{V}(\bar\psi\gamma_\mu\psi)(\bar\psi\gamma^\mu\psi)
	-\frac{1}{2}\alpha_{TV}(\bar\psi\vec\tau\gamma_\mu\psi)\cdot(\bar\psi\vec\tau\gamma^\mu\psi) \nonumber\\
	&& -\frac{1}{3}\beta_S(\bar\psi\psi)^3-\frac{1}{4}\gamma_S(\bar\psi\psi)^4
	-\frac{1}{4}\gamma_V[(\bar\psi\gamma_\mu\psi)(\bar\psi\gamma^\mu\psi)]^2 \nonumber\\
	&& -\frac{1}{2}\delta_S\partial_\nu(\bar\psi\psi)\partial^\nu(\bar\psi\psi)
	-\frac{1}{2}\delta_V\partial_\nu(\bar\psi\gamma_\mu\psi)\partial^\nu(\bar\psi\gamma^\mu\psi) \nonumber\\
	&& -\frac{1}{2}\delta_{TV}\partial_\nu(\bar\psi\vec\tau\gamma_\mu\psi)\cdot\partial^\nu(\bar\psi\vec\tau\gamma^\mu\psi)\nonumber\\
	&& -\frac{1}{4}F^{\mu\nu}F_{\mu\nu}  - e\bar\psi\gamma^\mu\frac{1-\tau_3}{2}\psi A_\mu,
\end{eqnarray}
which includes the free-nucleons term, the four-fermion point-coupling terms,
the higher-order terms which are responsible for the effects of medium dependence,
the gradient terms which are included to simulate the effects of finite range,
and the electromagnetic interaction terms.
The Dirac spinor field of the nucleon is denoted by $\psi$, and the nucleon mass is $m$.
$\vec\tau$ is the isospin Pauli matrix, and $\Gamma$ generally denotes the 4$\times$4 Dirac matrices
including $\gamma_\mu$, $\sigma_{\mu\nu}$ while
greek indices $\mu$ and $\nu$ run over the Minkowski indices 0, 1, 2, and 3.
$\alpha$, $\beta$, $\gamma$, and $\delta$ with subscripts $S$ (scalar), $V$ (vector), $TV$ (isovector) are coupling constants (adjustable parameters) in which $\alpha$ refers to the four-fermion term, $\beta$ and $\gamma$ respectively to the third- and fourth-order terms, and $\delta$ the derivative couplings.

Following the prescription in Ref.~\cite{Goodman1981} where the BCS limit of finite-temperature
Hartree-Fock Bogoliubov equations is derived, we obtain the finite-temperature CDFT + BCS equation.
The Dirac equation for single nucleons at finite-temperature reads
\begin{equation}\label{Eq:Dirac-PC}
	[\gamma_\mu(i\partial^\mu-V^\mu)-(m+S)]\psi_k=0,
\end{equation}
where $m$ is the nucleon mass.  $\psi_k(\bm{r})$ denotes the Dirac spinor field of a nucleon.
The scalar $S(\bm{r})$ and vector $V^\mu(\bm{r})$ potentials are
\begin{equation}\label{Eq:S}
	S(\bm{r})  =\alpha_S\rho_S+\beta_S\rho^2_S+\gamma_S\rho^3_S+\delta_S\triangle\rho_S,
\end{equation}
\begin{eqnarray}\label{Eq:V}
	V^\mu (\bm{r}) &=&\alpha_Vj^{\,\mu}_V +\gamma_V (j^{\,\mu}_V)^3 +\delta_V\triangle j^{\,\mu}_V\nonumber\\
	& &+\tau_3\alpha_{TV} \vec{j}^{\,\mu}_{TV}+ \tau_3\delta_{TV}\triangle \vec{j} ^{\,\mu}_{TV}+ e A^\mu
\end{eqnarray}
respectively.
The isoscalar density $\rho_S$, isoscalar current $j^{\,\mu}_V$ and isovector current $\vec{j}^{\,\mu}_{TV}$ are
represented by,
\begin{eqnarray}
	\rho_S (\bm{r})      &=& \sum \limits_{k} \bar\psi_k(\bm{r}) \psi_k(\bm{r}) [v_k^2 (1-2 f_k)+f_k],\label{Eq:dencur1} \\
	j^{\,\mu}_V (\bm{r})     &=& \sum \limits_{k} \bar\psi_k(\bm{r}) \gamma^\mu \psi_k(\bm{r}) [v_k^2 (1-2 f_k)+f_k], \label{Eq:dencur2}\\
	\vec{j}^{\,\mu}_{TV} (\bm{r}) &=& \sum \limits_{k} \bar\psi_k(\bm{r}) \vec{\tau} \gamma^\mu \psi_k(\bm{r}) [v_k^2 (1-2 f_k)+f_k].\label{Eq:dencur3}
\end{eqnarray}
where the thermal occupation probability of quasiparticle states $f_k$
is directly related to the temperature $T$ by $f_k=1/(1+e^{E_k/k_BT})$.
$E_k$ is the quasiparticle energy for single-particle (s.p.) state $k$,
$E_k = [(\epsilon_k-\lambda_q)^2 +(g_k \Delta_k)^2]^{\frac{1}{2}}$.
The smooth energy-dependent cutoff weight $g_k$ is determined to simulate the effect of finite range.
$\Delta_k$ is the pairing gap parameter, which satisfies the gap equation at finite temperature.
\begin{equation}
	\Delta_k = - \frac{1}{2} \sum_{k'>0} V^{pp}_{k\bar{k} k' \bar{k}'} \frac{\Delta_{k'}}{ E_{k'}} (1-2f_{k'}).
\end{equation}
In Eqs.~(\ref{Eq:dencur1})-(\ref{Eq:dencur3}),
the BCS occupation probabilities $v_k^2$ and associated $u_k^2=1-v_k^2$ are obtained by
\begin{eqnarray}\label{eq:occ}
	v_k^2   &=&\frac{1}{2} (1- \frac{\epsilon_k-\lambda_q}{E_k}), \\ 
	u_k^2   &=&\frac{1}{2} (1+ \frac{\epsilon_k-\lambda_q}{E_k}).
\end{eqnarray}

The particle number $N_q$ is restricted by $N_q= 2 \sum \limits_{k>0} [v_k^2 (1-2 f_k)+f_k]$.
Here we take the $\delta$ pairing force $V(\bm{r})=V_q\delta(\bm{r})$, where $V_q$ is the pairing strength parameter for neutrons or protons.
For the odd-$A$ nuclei, the unpaired nucleon is blocked automatically in the BCS approach.
For each iteration, the paired nucleons are filled to the single-particle levels from the bottom, then 
the first single-particle level above these levels is occupied by the unpaired nucleon with fixed $v^2=0.5$.
This specific level is picked out while other levels entering a small pairing loop yielding self-consistent $v_k$ and Fermi energy $\lambda$.

The internal binding energies $E$ at different shapes
can be obtained by applying constraints with quadrupole deformation $\beta_2$ and triaxial deformation $\gamma$ together.

The free energy for the system is $F=E-TS$ where the entropy $S$ is evaluated from
\begin{equation}\label{eq:S}
	S=-k_B \sum \limits_{k} [f_k{\rm ln}f_k +(1-f_k){\rm ln}(1-f_k)].
\end{equation}
For convenience, the temperature used is $k_BT$ in units of MeV and
the entropy used is $S/k_B$ and is unitless.
The free energy surface in the $(\beta_2,\gamma)$ plane is obtained by plotting
the free energy $E(\beta_2,\gamma)-TS(\beta_2,\gamma)$ on a mesh with equidistant $\beta_2$ and $\gamma$.

The specific heat is defined by the relation
\begin{equation}
	C_{\rm v}=\partial E^*/\partial T,
\end{equation}
where $E^*(T)=E(T)-E(T=0)$ is the internal excitation energy, and
$E(T)$ is the internal binding energy for the global minimum state in the free energy surface at certain temperature $T$.
More details can be found in Ref.~\cite{NST2024}.

\subsection{The DNS model}

Volkov et al.'s DNS concept pioneered the understanding of deep-inelastic heavy-ion collisions \cite{VOLKOV197893}. Non-equilibrium quantum-statistical mechanics has been instrumental in elucidating the dynamics of nucleon transfer in the DNS \cite{NORENBERG1974289,Norenberg1975}. The model was further refined in Lanzhou, incorporating the interplay of kinetic energy and impact parameters with nucleon rearrangement \cite{PhysRevC.76.044606}. It outlines a pathway involving capture, fusion, and survival of excited compound nuclei, pivotal for SHN synthesis \cite{FENG200650}.
Based on the DNS model, the evaporation residual cross-section of SHN is expressed as a sum over partial waves with angular momentum $J$ at the centre-of-mass energy $E_{\rm c.m.}$,
\begin{align}
	\sigma _{\rm ER}\left ( E_{\rm c.m.} \right ) = & \frac{\pi \hbar ^2}{2\mu E_{\rm c.m.}} \sum_{J=0}^{J_{\rm max}}(2J+1)T(E_{\rm c.m.},J)\nonumber \\ &\times P_{\rm CN}(E_{\rm c.m.},J)W_{\rm sur}(E_{\rm c.m.},J). 
\end{align}
Here, the $T(E_{\rm c.m.},J)$ is the probability of the collision system passing through the Coulomb barrier \cite{FENG200650}. The $P_{\rm CN}(E_{\rm c.m.},J)$ is the fusion probability \cite{PhysRevC.80.057601,PhysRevC.76.044606}. The $W_{\rm sur}(E_{\rm c.m.},J)$ is the probability of survival. Here, we take the maximal angular momentum as $J_{\rm max}$ = 100 $\hbar$.

The $T\left(E_{\mathrm{c} . \mathrm{m} .}, J\right)$ is derived by the Hill-Wheeler formula \cite{PhysRev.89.1102} with barrier distribution function \cite{PhysRevC.101.024610}, written as
\begin{eqnarray}\label{hwl}
	&T(E_{\mathrm{c.m.}},J)=\displaystyle\int \frac{f(B) \mathrm{d}B}{1+\exp\Big\{- \frac{2\pi [E_{\mathrm{c.m.}}-B-E^{\rm P-T}_{\rm rot}(J)]}{\hbar\omega(J)}\Big\}}.
\end{eqnarray}
Here, \(\hbar \omega(J)\) represents the width of the parabolic barrier $V$ at \(R_{\mathrm{B}}(J)\) position, given by  
\( \hbar \omega(J) = \hbar \sqrt{-({1}/{\mu}) (\partial^2 V/\partial R^2) }|_{R=R_{\mathrm{B}}(J)}. \)
$R_{\rm B} (J) $ is the position of the Coulomb barrier $B$. 
The reduced mass, denoted by \(\mu\), is defined as  
\( \mu = (m_{\text{P}} \cdot m_{\text{T}})/ (m_{\text{P}} + m_{\text{T}})\), where \(m_{\text{P}}\) and \(m_{\text{T}}\) are the masses of the projectile and target nuclei, respectively.
The $E^{\rm P-T}_{\rm rot}(J)$ is only the relative rotation energy of the projectile and target nuclei at $R_{\rm B} (J) $.
The interaction potential $V = V_{\rm C} + V_{\rm N}$, where the Coulomb potential $V_{\rm C}$ is evaluated by Wong's formula \cite{PhysRevLett.31.766}, and the nucleus-nucleus potential $V_{\rm N}$ is calculated by the double-folding method \cite{PhysRevC.80.057601,PhysRevC.76.044606, J.Mod.Phys.E5191(1996)}, using nucleon density distributions of the Fermi-type form with a surface diffuseness of $a=0.56$ fm. The Coulomb barrier distribution is taken as the asymmetric Gaussian formula \cite{PhysRevC.65.014607,Chen2017}.
\begin{eqnarray}\label{asg}
	f(B)=
	\left\{\begin{matrix}
		\frac{1}{N}\exp \left [-(\frac{B-B_{\rm m}}{\Delta_{1} } )\right ]\ \  B<B_{\rm m}\\
		& \\\frac{1}{N}\exp \left [-(\frac{B-B_{\rm m}}{\Delta_{2} } )\right ]\ \ B>B_{\rm m}.&
	\end{matrix}\right.
\end{eqnarray}
Here, $\bigtriangleup _2=\left ( B_0-B_{\rm S} \right )/2$, $\bigtriangleup_1=\bigtriangleup_2-2$ MeV, $B_{\rm m}=(B_0+B_{\rm s})/2$. The Coulomb barriers \( B_0 \) and \( B_{\rm s} \) correspond to distinct stages of a nuclear interaction: \( B_0 \) arises at the contact point in a side-by-side collision configuration, while \( B_{\rm s} \) occurs at the saddle point. The normalization constant satisfies \(\int f(B)  dB = 1\).  
For the deformed target-projectile system, the influence of different collision orientations on the capture probability is primarily reflected in the barrier distribution. The range of the barrier distribution is taken to lie between the barriers for the head-on (0° -- 0°) and side-on (90° -- 90°) configurations \cite{Zagrebaev2001,Zagrebaev2002}.

In the fusion stage, the formation probabilities of fragments P($Z_{1},N_{1},E_{1}$) are estimated by solving a set of master equations \cite{PhysRevC.76.044606,FENG200933,FENG201082}, 
\begin{eqnarray}
	&&\hspace{-8mm}\frac{d P(Z_1,N_1,E_1,t)}{d t} =  \nonumber \\ 
	&&\hspace{-8mm} \sum \limits_{Z'_1}W_{Z_1,N_1;Z'_1,N_1}(t)\times \nonumber \\ 
	&&\hspace{-8mm}\left[d_{Z_1,N_1}P(Z'_1,N_1,E'_1,t) - d_{Z'_1,N_1}P(Z_1,N_1,E_1,t) \right]  \nonumber \\ 
	&&\hspace{-8mm} +\sum \limits_{N'_1}W_{Z_1,N_1;Z_1,N'_1}(t) \times \nonumber \\ 
	&&\hspace{-8mm} \left[d_{Z_1,N_1}P(Z_1,N'_1,E'_1,t) - d_{Z_1,N'_1}P(Z_1,N_1,E_1,t) \right]. 
\end{eqnarray}
Here, the $W_{\rm Z_1,N_1,Z'_1,N_1}$ is the mean transition probability from the channel ($Z_1,N_1,E_1$) to ($Z'_1,N_1,E'_1$), and $d_{\rm Z_1,N_1}$ denotes the microscopic dimension corresponding to the macroscopic state ($Z_1,N_1,E_1$) \cite{Feng:2006zeg}.
The summation runs over all possible proton and neutron numbers accessible to the fragment $(Z'_1,N'_1)$; within the model, however, only one-nucleon transfer is allowed at a time, so the allowed values are restricted to  
$Z'_1 = Z_1 \pm 1$ and $N'_1 = N_1 \pm 1$. 
The $E_1$ is the local excitation energy $\varepsilon^*_1$ of fragment $(Z_1,N_1)$, generated by energy dissipation from the relative motion \cite{PhysRevC.27.590}. The sticking time is derived by the parametrization method of the classical deflection function \cite{LI1981107}.

The driving potential energy of the DNS is derived by
\begin{eqnarray}
	U_{\rm dr}(\{\alpha\},J) &=& B_1+B_2-\left[B_{\rm CN}+E^{\rm CN}_{\rm rot}(J) \right]+ \nonumber \\ 
	&& V(\alpha) + E^{\rm P-T}_{\rm rot}(J). 
\end{eqnarray}
Here, the $\{\alpha\}$ stands for $\{A_1,A_2,r,\beta_{1},\beta_{2},\theta _{1},\theta_{2}\}$, which are the mass numbers, surface distance, quadrupole deformations, and collision orientations of the colliding partners, respectively. $B_{\rm i}$ ($\rm i$ = 1, 2) and $B_{\rm CN}$ are the negative binding energies of the fragment $A_{\rm i}$ and the compound nucleus $A_{\rm CN}$, respectively, which are taken from TABLE \ref{tab1} and the 
FRDM \cite{PhysRevC.91.024310}. 
The $E^{\rm CN}_{\rm rot}(J)$ is the rotational energy of the compound nucleus.
The rotation energy $E^{\rm P-T}_{\rm rot}(J)$ here is calculated at the potential pocket.
The DNS model posits that all fragments capable of surpassing the Businaro-Gallone (BG) point can undergo fusion. Hence, the fusion probability can be written as 
\begin{eqnarray}
	P_{\rm CN}(E_{\rm c.m.},J)=\sum _{Z=1}^{Z _{\rm BG}}\sum_{N=1}^{N_{\rm BG}} P(Z,N,E',\tau_{\rm int}(J)).
\end{eqnarray}
Here $Z _{\rm BG}$ and $N_{\rm BG}$ are the proton number and neutron number of the BG point. The fusion probability is calculated not by considering the barrier distribution, but rather by using the potential energy at the bottom of the potential pocket, assuming a tip-to-tip collision.
$\tau_{\rm int} (J)$ is the sticking time with given angular momentum $J$.

Understanding of the mechanisms involved in the dynamic fusion stage of SHN synthesis remains incomplete, requiring further investigation. This stage can be described not only by the DNS mechanism but also by the shape evolution mechanism\cite{Denisov:2021tmc,Albertsson:2024tyb}, which is currently a hot topic of discussion. Additionally, the probability of compound nucleus formation can also be evaluated through various phenomenological or semi-phenomenological expressions. The DNS mechanism adopted in this paper can offer a clear physical picture, comprehensively account for key physical effects, and demonstrate excellent performance in predicting superheavy element synthesis.

The compound nuclei formed by all the nucleons transfer from projectile nuclei to target nuclei with certain excitation energies. 
The excited compound nuclei is extremely unstable which would be de-excited mainly by evaporating neutrons and $\gamma$-rays against fission. The survival probability of the $x$-neutron channels is given by a developed statistical evaporation model based on Weisskopf's evaporation theory \cite{Weisskopf1985,Chen2016,FENG201082}.
\begin{eqnarray}
	W^x_{\rm sur}(E_{\rm c.m.},J)=P(E_{\rm CN}^*,x,J)   \prod_{i=1}^{x}\frac{\Gamma _{\rm n}(E_i^*,J)}{\Gamma _{\rm tot}(E_i^*,J)}, 
\end{eqnarray}
where the $E_{\rm CN}^*$ and $J$ are the excitation energy and the spin of the excited nucleus, respectively. Here, the excitation energy of the compound nuclei is defined as $E_{\mathrm{CN}}^*=E_{\mathrm{c.m.}}+Q$. The total width $\Gamma_{\rm tot}$ is the sum of partial widths of neutron evaporation, $\gamma$-rays, and fission, as $\Gamma_{\rm tot}(E_i^*,J)=\Gamma_{\rm n}(E_i^*,J)+\Gamma_{\rm f}(E_i^*,J)+\Gamma_{\rm \gamma}(E_i^*,J)$. The excitation energy $E_i^*$ before evaporating the $i$-th particles is evaluated by
\begin{eqnarray}
	E_{i+1}^*=E_{i}^* - S_{\rm n}^i  - 2T_i
\end{eqnarray}
with the initial condition $E_1^*$=$E_{\rm CN}^*$. The $S_{\rm n} ^i$ is the separation energy of the $i$-th neutron. The nuclear temperature $T_i$ is defined by $T_i=\sqrt{E_{i}^*/a}$ where $a$ is the level density parameter. The decay width of the $\gamma$-ray are evaluated with the method in Ref.\cite{Snover1986,Chen2016}.
The $P(E_{\rm CN}^*,x,J)$ is the realization probability of evaporation channels \cite{Jackson1956}.  

The neutron decay width is evaluated with the Weisskopf evaporation theory~\cite{PhysRevC.68.014616} as
\begin{eqnarray}
	&&\Gamma_{\rm n}(E^*,J)=(2s+1)\frac{m_{\rm n}}{\pi ^2\hbar ^2\rho (E^*,J)}\nonumber\\&& \times \int_{0}^{\varrho -\frac{1}{a} }\varepsilon \rho (\varrho +\delta-E^{\rm CN}_{\rm rot}-\varepsilon ,J) \sigma _{\rm inv}d\varepsilon.
\end{eqnarray}
Here, we set $E^*-B_{\rm n}-\delta-\delta_\mathrm{n}$ to the term $\varrho$. $s$, $m_{\rm n}$ and $B_{\rm n}$ are the spin, mass and binding energy of the neutron, respectively. The pairing correction energy $\delta$ is set to be $12/\sqrt{A}$, 0, $-12/\sqrt{A}$ for even-even, even-odd and odd-odd nuclei, respectively. $\delta_{\rm n}$ is the pairing correction for neutron evaporation. The $\sigma_{\rm inv}=\pi R_{\rm n}^{2}T(n) $ is the cross section for inverse process.  $R_{\rm n}$ is the radius of the neutron-nucleus interaction \cite{Weisskopf1985}.

The fission width $\Gamma_{\rm f}(E^*,J)$ is calculated by the modified Bohr-Wheeler formula  as in Ref.\cite{PhysRevC.80.057601,PhysRevC.76.044606}. 
An alternative method for calculating the fission width exists, which also incorporates the excitation-energy-dependent fission barrier \cite{Denisov2018}.
We set $ E^*-B_{\rm f}(E^*)-E^{\rm CN}_{\rm rot}-\delta -\delta _{\rm f}$ to the term $ \kappa $.
\begin{eqnarray}
	&&\Gamma_{\rm f}(E^*,J)=\frac{1}{2\pi \rho_{\rm f} (E^*,J)}\int_{0}^{\kappa -\frac{1}{\alpha _{\rm f}} }\nonumber \\&&  
	\frac{\rho _{\rm f}(\kappa -\varepsilon+\delta ,J)d\varepsilon }{1+\rm exp\left [ -2\pi (\kappa -\varepsilon+\delta+\delta_{\rm f} )/\hbar \omega  \right ] } 
\end{eqnarray}
For heavy fragments, the curvature of the fission barrier at the saddle point is usually taken as $\hbar\omega=$ 2.2 MeV \cite{artza05}, and $\delta_{\rm f}$ is the pairing correction for the fission barrier. 
The back-shift Fermi gas energy level density can be expressed as
\begin{eqnarray}
	\rho (E^*,J)=K_{\rm coll}\times \frac{2J+1}{24\sqrt{2}\sigma ^3a^{1/4}(E^*-\delta )^{5/4} } \nonumber \\ \times  \exp\left [ 2\sqrt{a(E^*-\delta )}-\frac{(J+1/2)^2}{2\sigma ^2}   \right ]  
\end{eqnarray}
with $\sigma^2 = 6\bar{m}^2\sqrt{a(E^*-\delta)}/\pi^2$ and $\bar{m}\approx0.24A^{2/3}$. The $K_{\rm coll}$ is the collective enhancement factor which contains the rotational and vibration effects. The ground state level density parameter $a$ takes the asymptotic value $\tilde{a}$, $a_{\rm f}$=1.1$a$ for the fission level density parameter at saddle point.

In this calculation, the $J$-dependent fission barrier is:
\begin{eqnarray}
	\label{Bf}
	B_{\rm f}(E^*,J)=B_{\rm f} (E^*)+ 
	(\frac{\hbar^2}{2\zeta_{\rm sd}} - \frac{\hbar^2}{2\zeta_{\rm gs}})J(J+1),
\end{eqnarray}
where the $B_{\rm f}(E^*)$ is derived by the finite-temperature CDFT calculations using PC-PK1 energy density functional. 
$\zeta_{\rm gs}$ and $\zeta_{\rm sd}$ are the moments of inertia for the ground state and the saddle point, respectively, of a compound nucleus with spin $J$ \cite{FENG200650}.

		\section{Results and discussion}\label{sec3}

Firstly, the thermal properties of the compound nuclei are obtained by the finite-temperature CDFT. The free energy surfaces at different temperatures including saddles and minima are presented. 
Subsequently, the behaviors of excitation energy $E^{*}$ and entropy $S$ with temperature $T$ are discussed, and the level density parameters $a$ is extracted based on these quantities.
Inserting $E_{\mathrm{sh}}$ calculated by the Strutinsky shell correction method, the asymptotic level density parameter $\tilde{a}$ and the shell damping factor $E_{\mathrm{D}}$ are fitted by the phenomenological formula to the level density parameter $a$.
Other quantities like single neutron separation energy $S_{\mathrm{n}}$ and fission barrier $B_{\mathrm{f}}(E^*)$ are also obtained in the CDFT.
Finally, these input physical quantities are used in the DNS model. The corresponding excitation functions 
of several reaction systems are compared with experimental data. And the synthesis cross section for element 120 is predicted.

\begin{figure*}[htbp]
	\includegraphics[width=1.\linewidth]{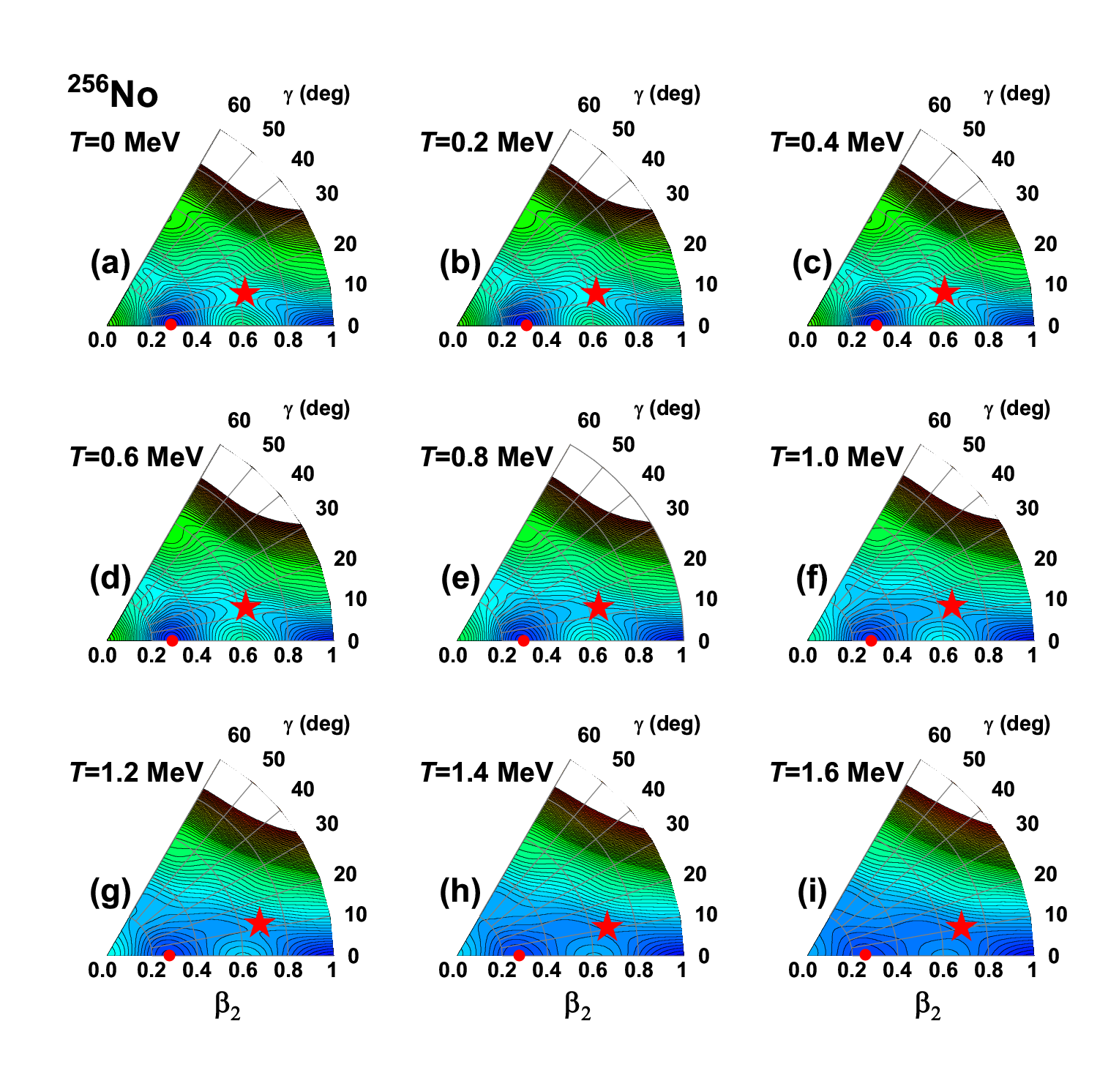}
	\caption{\label{fig1} Panels (a) to (i) depict the free energy surfaces in the ($\beta_2$,$\gamma$) plane at different temperatures ($T=0$, 0.2, 0.4, 0.6, 0.8, 1.0, 1.2, 1.4, and 1.6 MeV) for compound nucleus $^{256}\text{No}$ obtained by the finite-temperature CDFT calculations using PC-PK1 energy density functional. The energy separation between contour lines is 0.5 MeV. The global minimum and saddle points are represented by the red dots and stars, respectively.}
\end{figure*}

To obtain the relevant physical quantities as input parameters to the DNS model, large scale calculations have been performed for compound nuclei $^{248-256}$No, $^{278-292}$Fl, and $^{290-302}$120 in the $(\beta_2,\gamma)$ plane using the finite-temperature CDFT using PC-PK1 energy density functional. 
Taking $^{256}$No as an example, Fig. \ref{fig1} shows the evolution of free energy surfaces in the $(\beta_2,\gamma)$ plane as a function of temperature $T$ in the range 0$-$1.6 MeV with a step of 0.2 MeV. 
It is clear that the deformation positions $(\beta_2,\gamma)$ of the global minima and saddle points 
represented by the dots and stars respectively remain nearly where they are as in the zero temperature.
The energy difference between the global minimum and saddle point is the fission barrier height $B_{\mathrm{f}}$, counting by the number of contour lines in between visually in this figure. 
For $T\le$ 0.4 MeV in Fig. \ref{fig1} (a) to (c), the fission barrier remains unchanged,
while for higher temperature it decrease exponentially.
The evolution of the free energy surfaces of compound nuclei $^{248-255}$No with temperature is similar to that of $^{256}$No.

\begin{figure}[htbp]
	\includegraphics[width=1.\linewidth]{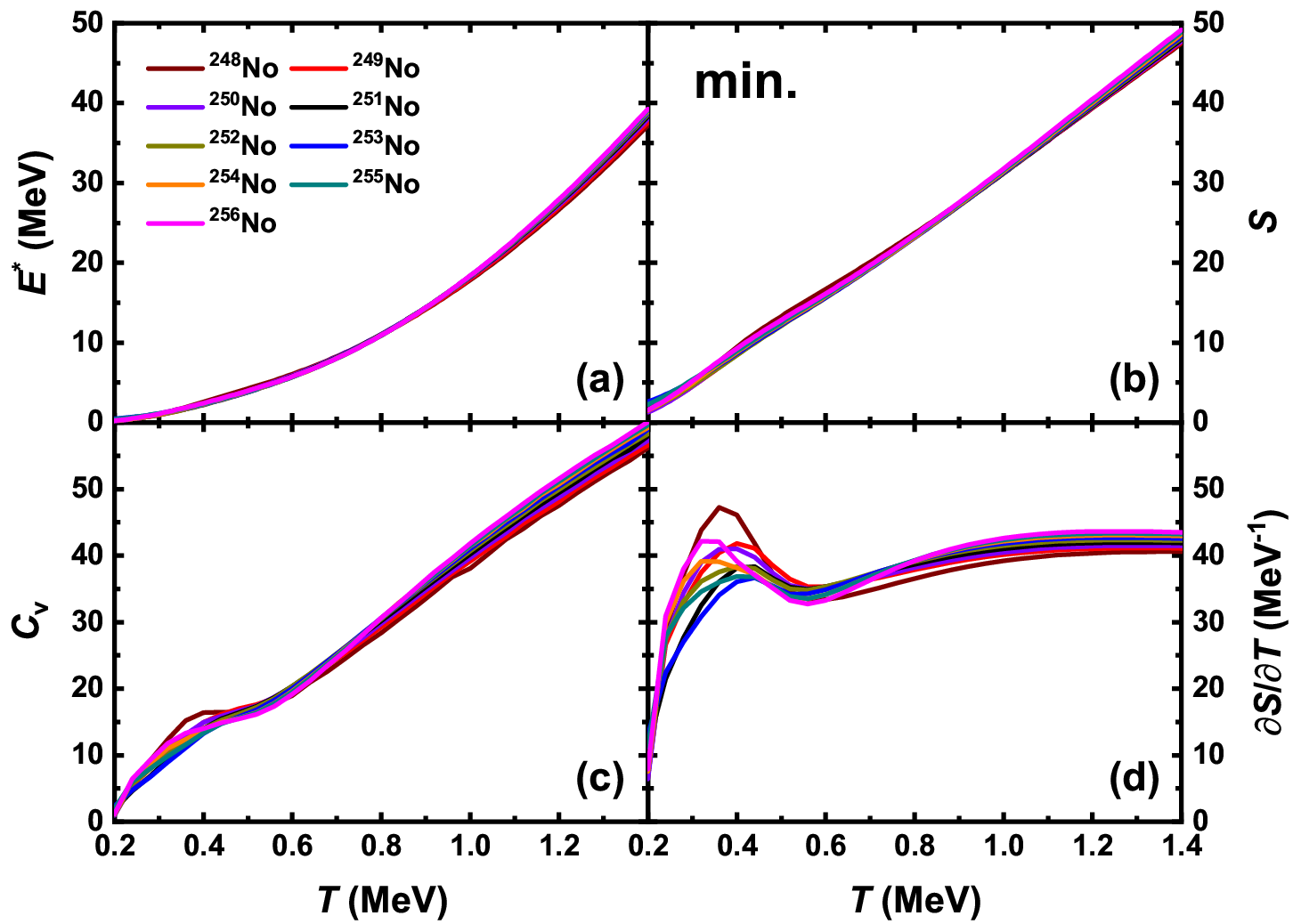}
	\caption{\label{fig2} Panels (a) to (d) depict the excitation energy $E^{*}$, entropy $S$, specific heat $C_\mathrm{v}$, and $\partial S/\partial T$ as functions of temperature $T$ for global minimum of compound nuclei $^{248-256}$No obtained by the finite-temperature CDFT calculations using the PC-PK1 energy density functional. Solid lines of different colors represent different compound nuclei.}
\end{figure}

Based on the free energy surfaces, Fig. \ref{fig2} shows the behaviors of the excitation energy $E^{*}$, entropy $S$, specific heat $C_{\rm v}$ and partial derivative of entropy with respect to temperature $\partial S/\partial T$ for compound nuclei $^{248-256}$No at the global minimum.
For $C_\mathrm{v}$ in Fig. \ref{fig2}(c), this is basically a straight line. 
If you look into it closer, there are two segments $T <$ 0.4 MeV and $T >$ 0.5 MeV connecting together.
This is the critical temperature due to the phase transition. More details can be found in Ref. \cite{ZN2017,ZN2018}. 
Since the relation $C_{\rm v}=\partial E^*/\partial T$, the curve in Fig. \ref{fig2}(a) is actually a quadratic parabola.
This is consistent with the simple formula $E^{*} = aT^{2}$ where $a$ is the level density parameter in the Fermi-gas model \cite{Bethe1937}.
For $\partial S/\partial T$ in Fig. \ref{fig2}(d), the curve is generally a constant, 
especially for the high temperatures. 
For low temperatures, it increases quickly and reaches a peak.
The discontinuities between $T = 0.4$ and $0.5$ MeV are in consistency with Fig. \ref{fig2}(c) 
where the phase transition occurs.
Due to the relation between $S$ and $\partial S/\partial T$ shown in Fig. \ref{fig2}(b) and Fig. \ref{fig2}(d), 
the entropy $S$ increases proportionally with a constant slope for $T >$ 0.5 MeV.
Actually this phenomenon is well described by the classical relationship $S = 2aT$ \cite{Bethe1937}.
Note the constant height at high temperature shown in Fig. \ref{fig2}(d) is just twice 
the level density parameter $a$.

\begin{figure}[htbp]
	\includegraphics[width=1.\linewidth]{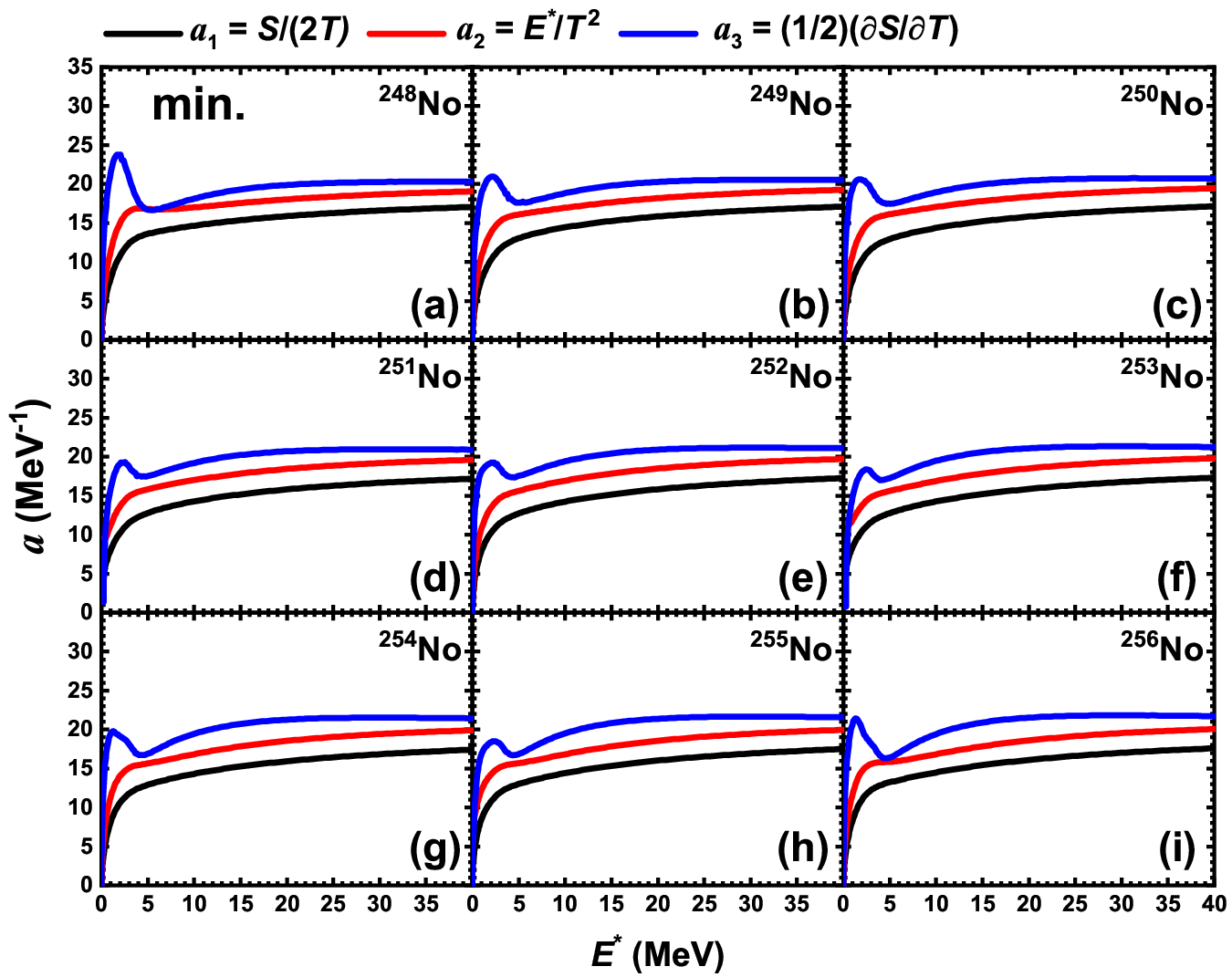}
	\caption{\label{fig3} Panels (a) to (i) depict the level density parameters $a$ as functions of the excitation energy $E^{*}$ at the global minimum for compound nuclei $^{248-256}$No extracted using $a_{1}=S/(2T)$ (black lines), $a_{2}=E^{*}/T^{2}$ (red lines), and $a_{3}=(1/2)(\partial S/\partial T)$ (blue lines).}
\end{figure}

From the relations $E^{*} = aT^{2}$ and $S = 2aT$ \cite{Bethe1937}, the level density parameters $a$ at the global minimum for compound nuclei $^{248-256}$No can be extracted from Fig. \ref{fig2}(a) and \ref{fig2}(b) using at least three ways $a_{1}=S/(2T)$, $a_{2}=E^{*}/T^{2}$, and $a_{3}=(1/2)(\partial S/\partial T)$. Fig. \ref{fig3} illustrates the level density parameters $a_{1}$, $a_{2}$, and $a_{3}$ as functions of the excitation energy $E^{*}$. 
Since the excitation energy curves in Fig. \ref{fig2}(a) and the entropy curves in Fig. \ref{fig2}(b)  coincide for different compound nuclei, it is expected that the level density parameter $a$ 
extracted by the same method for these compound nuclei are very similar between subfigures in Fig. \ref{fig3}.
In this figure, $a_3$ just copies the behavior shown in Fig. \ref{fig2}(d), while $a_1$ and $a_2$ maintain a monotonically increasing trend with excitation energy $E^{*}$. As the excitation energy increases from zero, the level density parameters $a_{1}$, $a_{2}$, and $a_{3}$ increase rapidly,
then become stable for $E^{*} >$ 4 MeV corresponding to the phase transition discussed before.
This correspondence can be found in Fig. \ref{fig2}(a) directly.
For $E^{*} >$ 15 MeV, all three exhibit identical asymptotic trends going paralleled.
Ultimately, they approach different asymptotic values at high excitation energy respectively. Among them, the asymptotic value of $a_{3}$ is the largest and the asymptotic value of $a_{1}$ is the smallest.

\begin{figure}[htbp]
	\includegraphics[width=0.95\linewidth]{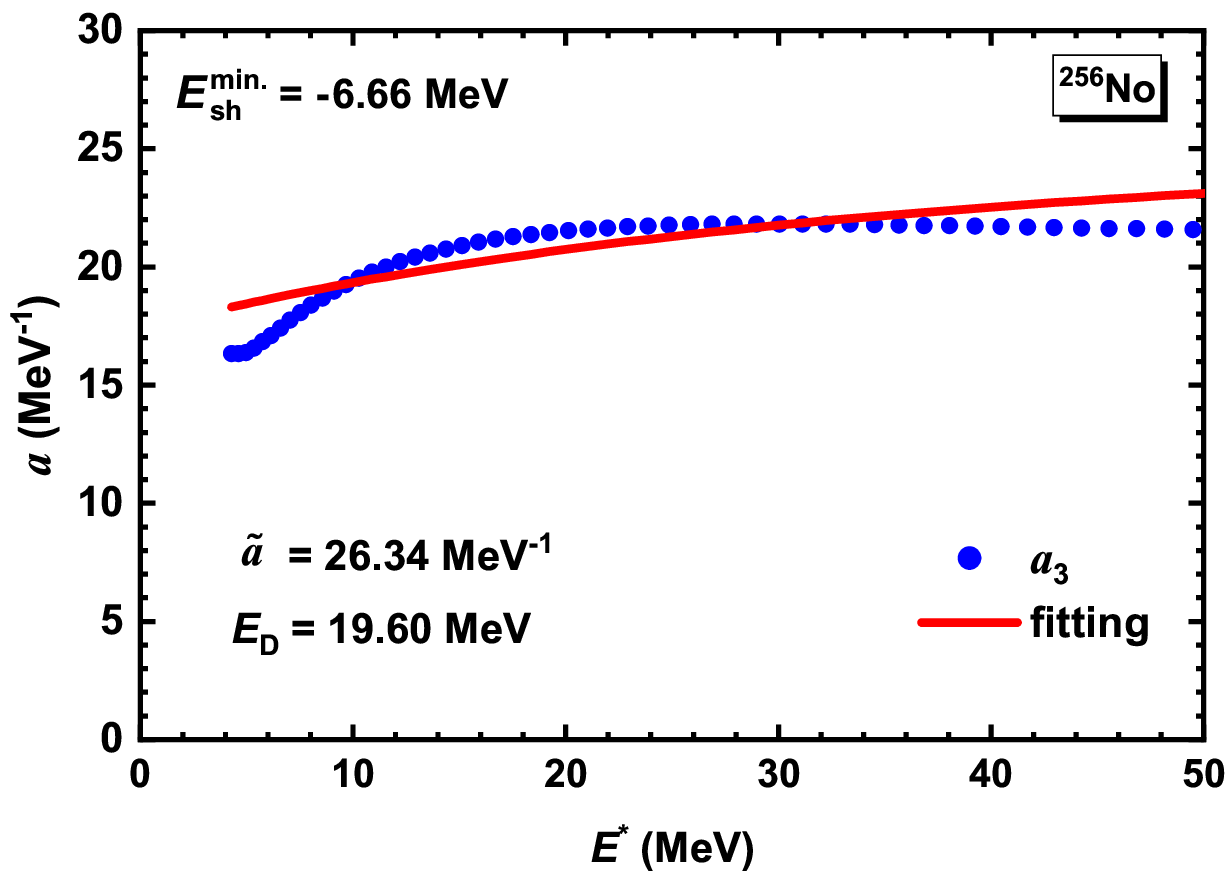}
	\caption{\label{fig4}
		The fitting result for the level density parameter $a_{3}$ at the global minimum of compound nucleus $^{256}$No using phenomenological expression $a(E^*) = \tilde{a} \left \{ 1 + E_{\mathrm{sh}} \left [1 - \exp(-E^*/E_{\mathrm{D}})\right ] /E^* \right \}$ \cite{Ignatyuk1975}. The shell correction energy $E_{\mathrm{sh}}$ is calculated from CDFT by using the Strutinsky shell correction method. 
	}
\end{figure}

In Fig. \ref{fig3}, the level density parameter $a$ approaches an asymptotic value $\tilde{a}$ when the excitation energy is sufficiently large, and the rate it approaches $\tilde{a}$ is determined by the shell damping factors $E_{\mathrm{D}}$.
To obtain the asymptotic level density parameters $\tilde{a}$ and the shell damping factors $E_{\mathrm{D}}$, the extracted level density parameters are fitted by the least-squares method using phenomenological expression $a(E^*) = \tilde{a} \left \{ 1 + E_{\mathrm{sh}} \left [1 - \exp(-E^*/E_{\mathrm{D}})\right ] /E^* \right \}$  \cite{Ignatyuk1975}.
The shell correction energy $E_{\mathrm{sh}}$ is obtained from CDFT by using the Strutinsky shell correction method together with the order of Gauss-Hermite polynomials is set to $p = 2M = 6$ ,
and the smoothing parameter $\gamma=1.3 \hbar \omega_0$ with $\hbar \omega_0 = 41 A^{-1/3} (1\pm \frac{1}{3}\frac{N-Z}{A})$ MeV where the plus (minus) sign holds for neutrons (protons)~\cite{cpc46.2022}.
Fig. \ref{fig4} presents the fitting result 
for the level density parameter $a_{3}$ at the global minimum of compound nucleus $^{256}$No.
During the fitting process, we truncate the data with excitation energy below 4 MeV, which corresponds to the critical temperature shown in Fig. \ref{fig2}. The fitting yields the asymptotic level density parameter $\tilde{a}=26.34$ MeV$^{-1}$ and shell damping factor $E_{\mathrm{D}}=19.60$ MeV. Using the same method, the $\tilde{a}$ and $E_{\mathrm{D}}$ for compound nuclei $^{248-255}$No and other nuclei can also be obtained.

\begin{figure}[htbp]
	\includegraphics[width=1\linewidth]{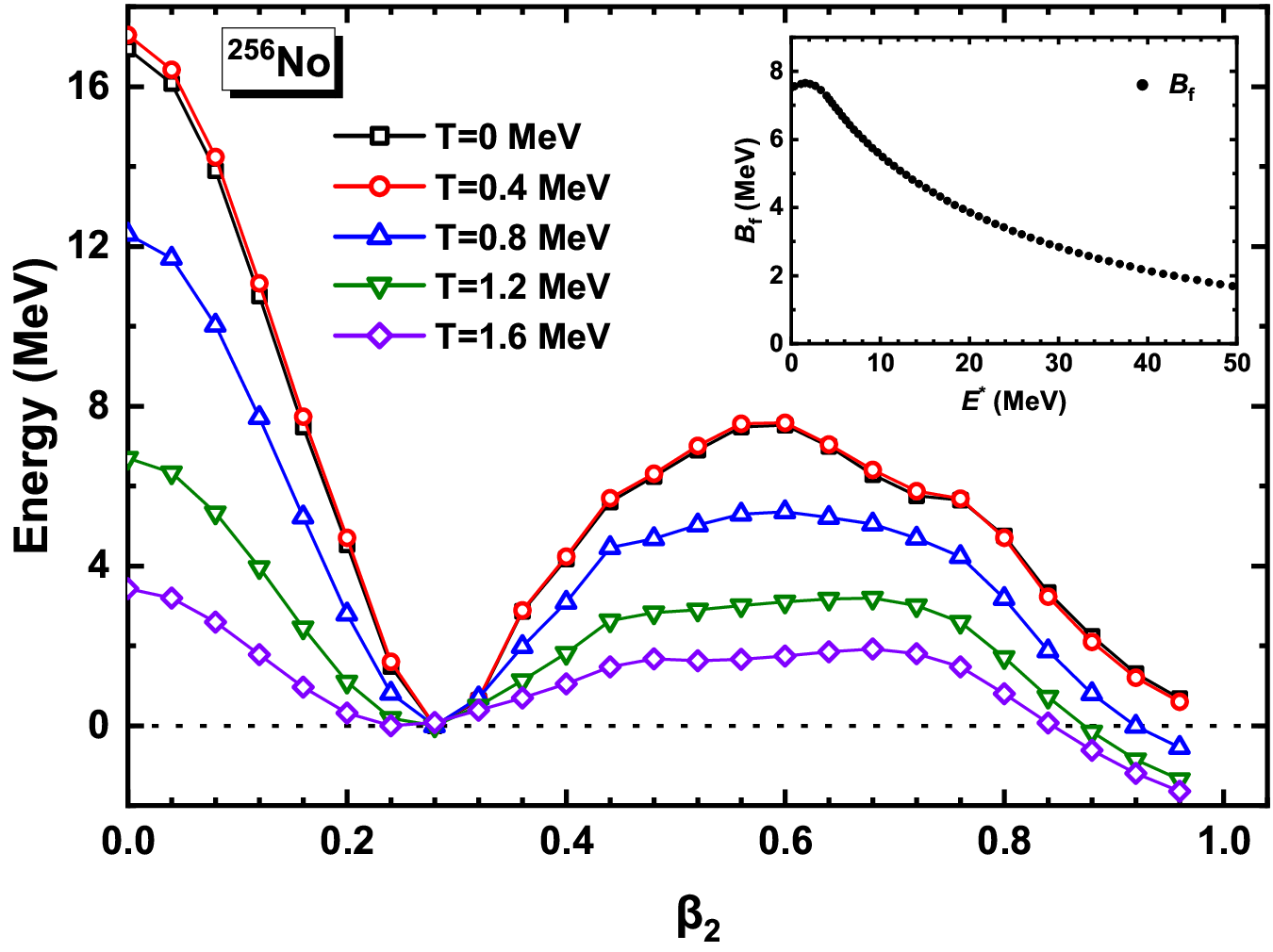}
	\caption{\label{fig5}
		Potential energy curves (PECs) for the compound nucleus $^{256}$No as functions of $\beta_2$ at different temperatures. The energy is normalized to zero at the ground-state minimum. The inset depicts the fission barrier $B_{\mathrm{f}}$ as a function of the excitation energy $E^{*}$ for the compound nucleus $^{256}$No.
	}
\end{figure}
To more intuitively demonstrate the effect of temperature on the fission barrier, we present the potential energy curves (PECs) for the compound nucleus $^{256}$No at different temperatures in Fig. \ref{fig5} , based on the results in Fig. \ref{fig1}.
The minimum point near $\beta_2 = 0.28$ corresponds to the ground state, while the maximum point near $\beta_2 = 0.60$ represents the saddle point. The energy difference between them is defined as the fission barrier $B_{\rm f}$.
It can be clearly seen that the fission barrier remains nearly constant for temperatures $T \leq 0.4$ MeV, and it decreases gradually with further increases in temperature.
The fission barrier at different excitation energies for the compound nucleus $^{256}$No is depicted in the inset of Fig. \ref{fig5}.
The fission barrier $B_{\mathrm{f}}(E^*)$ obtained via finite-temperature CDFT is employed in Eq.~(\ref{Bf}).
It can be observed that the fission barrier remains nearly constant for excitation energy below 4 MeV. As the excitation energy further increases, the fission barrier decreases exponentially.
The logarithm slope of the fission barrier is related to $E_{\mathrm{D}}$, indicating another method to extract the shell damping factor.
This will be discussed more deeply in the future.
In current work, $E_{\mathrm{D}}$ is obtained from the level density parameter. 


The single neutron separation energies $S_{\mathrm{n}}$, fission barriers $B_{\mathrm{f}}$, shell correction energies $E_{\mathrm{sh}}$, as well as the asymptotic level density parameters $\tilde{a}$ and shell damping factors $E_{\mathrm{D}}$ obtained using finite-temperature CDFT for compound nuclei $^{248-256}$No, $^{278-292}$Fl, and $^{290-302}$120 are given in TABLE \ref{tab1}.
The fitted  shell damping factors $E_{\mathrm{D}}$ are not used further and are only presented for reference.

\begin{table*}[htbp]
	\renewcommand{\arraystretch}{1.2}
	\centering
	\setlength{\tabcolsep}{3pt}
	\caption{\label{tab1} The single neutron separation energies $S_{\mathrm{n}}$, fission barriers $B_{\mathrm{f}}$, shell correction energies $E_{\mathrm{sh}}$, as well as the asymptotic level density parameters $\tilde{a}$ and shell damping factors $E_{\mathrm{D}}$ extracted using the three ways obtained by finite-temperature CDFT for compound nuclei $^{248-256}$No, $^{278-292}$Fl, and $^{290-302}$120. }
	\begin{tabular}{ccccccccccc}
		\toprule
		&  &  & \multicolumn{2}{c}{$E_{\mathrm{sh}}$} & \multicolumn{2}{c}{$a_{1}=S/(2T)$} &\multicolumn{2}{c}{$a_{2}=E^{*}/T^{2}$}  &\multicolumn{2}{c}{$a_{3}=(1/2)(\partial S/\partial T)$} \\
		\cmidrule(r){4-5}
		\cmidrule(r){6-7} 
		\cmidrule(r){8-9} 
		\cmidrule(r){10-11}
		& & & \multicolumn{1}{c}{(minimum)} & \multicolumn{1}{c}{(saddle)} &\multicolumn{2}{c}{(minimum)}  & \multicolumn{2}{c}{(minimum)} &  \multicolumn{2}{c}{(minimum)} \\
		Nucleus & $S_{\mathrm{n}}$ & $B_{\mathrm{f}}$ & $E_{\mathrm{sh}}^{\mathrm{min.}}$ & $E_{\mathrm{sh}}^{\mathrm{sd}}$ & $\tilde{a}$ & $E_{\mathrm{D}}$  & $\tilde{a}$ & $E_{\mathrm{D}}$ & $\tilde{a}$ & $E_{\mathrm{D}}$  \\
		& (MeV) & (MeV) & (MeV) & (MeV) & (MeV$^{-1}$) & (MeV)  & (MeV$^{-1}$) & (MeV)  & (MeV$^{-1}$) & (MeV)   \\
		\hline
		$^{248}$No	&	8.02 	&	6.59 	&	-5.24 	&	2.07 	&	19.10 	&	17.48 	&	21.72 	&	18.91 	&	23.35 	&	18.57 	 	\\
		$^{249}$No	&	7.73 	&	6.89 	&	-5.82 	&	2.13 	&	19.45 	&	17.47 	&	22.30 	&	18.87 	&	24.37 	&	19.93 	 	\\
		$^{250}$No	&	7.87 	&	7.38 	&	-6.76 	&	1.58 	&	20.16 	&	18.16 	&	23.34 	&	19.44 	&	25.45 	&	20.38 	 	\\
		$^{251}$No	&	7.44 	&	7.62 	&	-6.80 	&	2.45 	&	20.22 	&	18.19 	&	23.44 	&	19.30 	&	25.75 	&	20.99 	 	\\
		$^{252}$No	&	7.22 	&	7.73 	&	-7.08 	&	2.71 	&	20.44 	&	18.09 	&	23.75 	&	19.14 	&	26.27 	&	20.79 	 	\\
		$^{253}$No	&	7.02 	&	7.82 	&	-7.03 	&	2.29 	&	20.45 	&	17.99 	&	23.74 	&	18.94 	&	26.38 	&	20.61 	 	\\
		$^{254}$No	&	6.82 	&	7.67 	&	-7.17 	&	1.71 	&	20.66 	&	18.11 	&	23.92 	&	18.89 	&	26.65 	&	20.34 	 	\\
		$^{255}$No	&	6.60 	&	7.57 	&	-6.79 	&	1.82 	&	20.46 	&	17.81 	&	23.62 	&	18.44 	&	26.39 	&	20.03 	 	\\
		$^{256}$No	&	6.56 	&	7.52 	&	-6.66 	&	1.72 	&	20.48 	&	17.72 	&	24.23 	&	16.20 	&	26.34 	&	19.60 	 	\\
		$^{278}$Fl	&	7.52 	&	5.72 	&	-8.64 	&	1.49 	&	23.36 	&	19.08 	&	27.08 	&	19.68 	&	30.40 	&	21.68 	 	\\
		$^{279}$Fl	&	7.34 	&	5.54 	&	-8.26 	&	0.84 	&	23.11 	&	18.79 	&	26.74 	&	19.24 	&	30.10 	&	21.46 	 	\\
		$^{280}$Fl	&	7.34 	&	5.49 	&	-8.46 	&	0.75 	&	23.38 	&	18.82 	&	27.12 	&	19.29 	&	30.61 	&	21.72 	 	\\
		$^{281}$Fl	&	7.26 	&	5.49 	&	-8.65 	&	0.70 	&	23.63 	&	18.87 	&	27.57 	&	19.45 	&	31.08 	&	22.24 	 	\\
		$^{282}$Fl	&	7.21 	&	5.58 	&	-9.19 	&	0.56 	&	24.19 	&	19.08 	&	28.43 	&	19.75 	&	32.06 	&	22.92 	 	\\
		$^{283}$Fl	&	7.03 	&	5.89 	&	-9.28 	&	-0.85 	&	24.33 	&	19.02 	&	28.76 	&	19.76 	&	32.71 	&	23.03 	 	\\
		$^{284}$Fl	&	6.87 	&	6.10 	&	-9.59 	&	-0.32 	&	24.76 	&	19.13 	&	29.40 	&	19.96 	&	33.15 	&	24.31 	 	\\
		$^{285}$Fl	&	6.69 	&	6.41 	&	-9.62 	&	0.49 	&	24.93 	&	19.09 	&	29.69 	&	20.05 	&	33.25 	&	24.98 	 	\\
		$^{286}$Fl	&	6.55 	&	6.61 	&	-9.76 	&	1.09 	&	25.25 	&	19.25 	&	30.17 	&	20.40 	&	33.68 	&	25.60 	 	\\
		$^{287}$Fl	&	6.35 	&	6.76 	&	-9.75 	&	2.38 	&	24.96 	&	20.86 	&	29.13 	&	24.62 	&	33.67 	&	25.83 	 	\\
		$^{288}$Fl	&	6.34 	&	6.86 	&	-9.76 	&	2.78 	&	25.18 	&	21.20 	&	29.43 	&	25.80 	&	33.65 	&	26.34 	 	\\
		$^{289}$Fl	&	5.71 	&	6.39 	&	-8.40 	&	2.90 	&	24.44 	&	20.43 	&	28.10 	&	25.14 	&	31.87 	&	26.09 	 	\\
		$^{290}$Fl	&	5.81 	&	6.02 	&	-7.63 	&	2.91 	&	24.33 	&	21.80 	&	27.72 	&	27.22 	&	30.56 	&	26.70 	 	\\
		$^{291}$Fl	&	6.27 	&	6.22 	&	-4.17 	&	2.75 	&	22.10 	&	15.11 	&	25.29 	&	24.18 	&	27.39 	&	26.36 	 	\\
		$^{292}$Fl	&	6.25 	&	6.46 	&	-4.56 	&	2.73 	&	22.37 	&	16.09 	&	25.68 	&	24.89 	&	27.96 	&	26.07 	 	\\
		$^{290}$120	&	8.43 	&	4.77 	&	-5.60 	&	1.15 	&	22.33 	&	17.53 	&	25.66 	&	20.04 	&	27.25 	&	22.43 	 	\\
		$^{291}$120	&	8.15 	&	5.34 	&	-6.29 	&	1.78 	&	22.81 	&	17.80 	&	26.41 	&	20.12 	&	28.37 	&	23.35 	 	\\
		$^{292}$120	&	8.19 	&	6.02 	&	-7.30 	&	1.04 	&	23.63 	&	18.64 	&	27.59 	&	20.77 	&	29.57 	&	23.40 	 	\\
		$^{293}$120	&	7.67 	&	6.12 	&	-6.87 	&	-0.60 	&	23.35 	&	18.23 	&	27.19 	&	20.34 	&	29.23 	&	23.07 	 	\\
		$^{294}$120	&	7.71 	&	6.68 	&	-7.13 	&	1.05 	&	23.64 	&	18.41 	&	27.85 	&	21.11 	&	29.62 	&	23.41 	 	\\
		$^{295}$120	&	7.13 	&	6.50 	&	-6.66 	&	0.60 	&	23.32 	&	17.95 	&	27.03 	&	19.88 	&	29.28 	&	22.70 	 	\\
		$^{296}$120	&	7.32 	&	6.70 	&	-8.34 	&	1.03 	&	24.69 	&	19.16 	&	28.75 	&	20.48 	&	31.53 	&	22.90 	 	\\
		$^{297}$120	&	7.15 	&	6.78 	&	-8.45 	&	1.33 	&	24.86 	&	19.16 	&	29.03 	&	20.52 	&	31.84 	&	22.84 	 	\\
		$^{298}$120	&	7.02 	&	6.76 	&	-8.66 	&	0.81 	&	25.15 	&	19.31 	&	29.41 	&	20.69 	&	32.24 	&	22.78 	 	\\
		$^{299}$120	&	6.85 	&	6.65 	&	-8.60 	&	-0.29 	&	25.04 	&	18.99 	&	29.27 	&	20.00 	&	32.56 	&	22.55 	 	\\
		$^{300}$120	&	7.07 	&	6.97 	&	-8.62 	&	-0.25 	&	25.04 	&	18.79 	&	29.33 	&	19.68 	&	32.89 	&	22.52 	 	\\
		$^{301}$120	&	6.65 	&	7.08 	&	-8.69 	&	-0.70 	&	25.23 	&	18.71 	&	29.75 	&	19.96 	&	33.30 	&	23.52 	 	\\
		$^{302}$120	&	6.71 	&	7.44 	&	-8.99 	&	-0.73 	&	25.66 	&	18.99 	&	30.47 	&	20.47 	&	33.79 	&	24.09 	 	\\
		
		\bottomrule
	\end{tabular}
\end{table*}

\begin{figure}[htbp]
	\includegraphics[width=1.\linewidth]{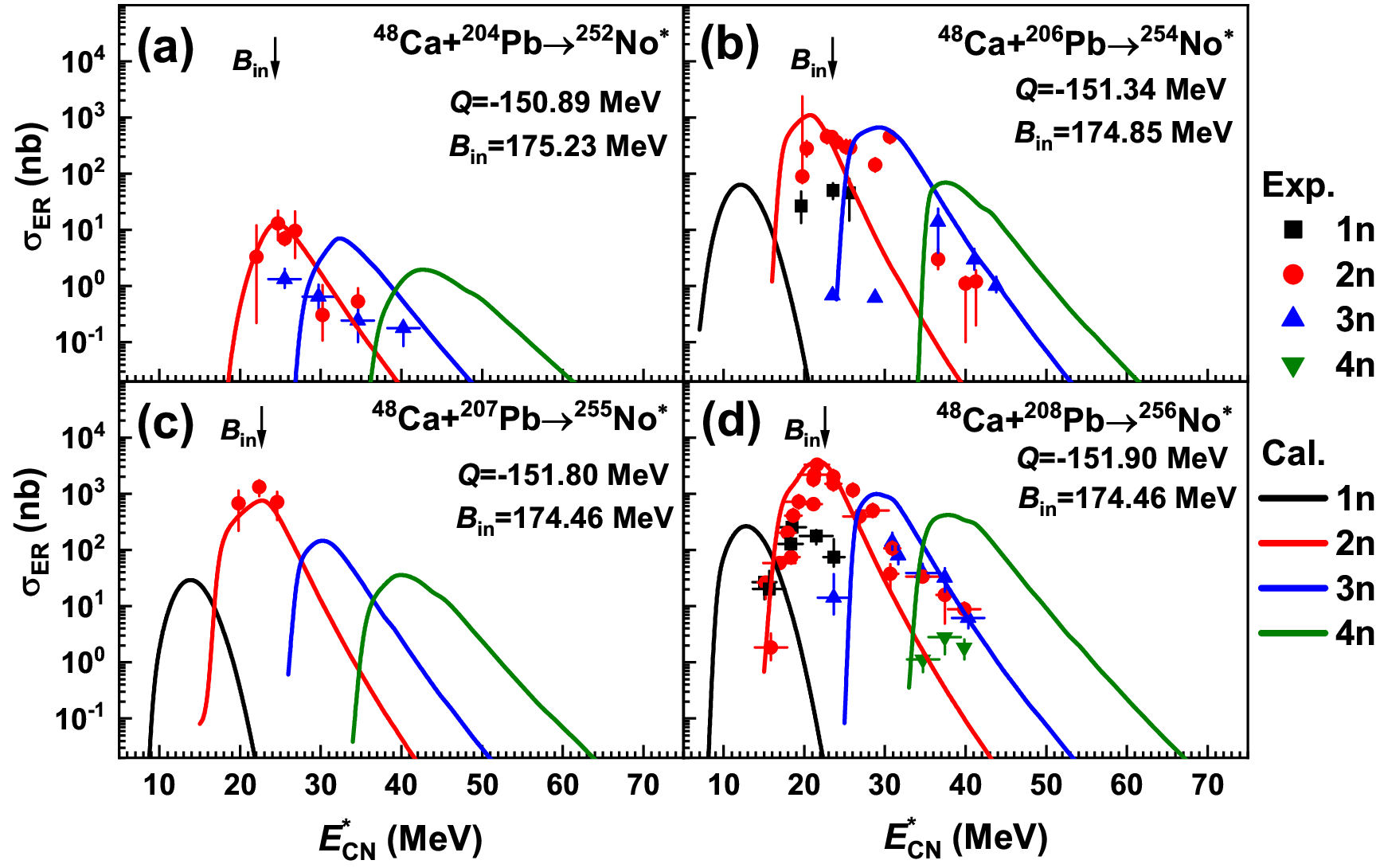}
	\caption{\label{fig6}
		Panels (a) to (d) depict the calculated excitation functions for Evaporation Residue Cross-Sections (ERCS) resulting from cold-fusion reactions of projectile nuclei $^{48}$Ca with target nuclei $^{204,206-208}$Pb, juxtaposed with experimental data indicated by error bars \cite{GAGGELER1989561,belozerov2003spontaneous,PhysRevC.69.054607}.
		The calculated ERCS for the 1n-, 2n-, 3n-, and 4n-channels are represented by black, red, blue, olive solid lines, respectively. Correspondingly, experimental data for the 1n-, 2n-, 3n-, and 4n-channels are marked by black solid squares, red solid circles, blue solid up-triangles, and olive solid down-triangles, respectively. The black arrows in the figure signify the Bass potential $B_{\rm in}$.
	}
\end{figure}

\begin{figure}[htbp]
	\includegraphics[width=1.\linewidth]{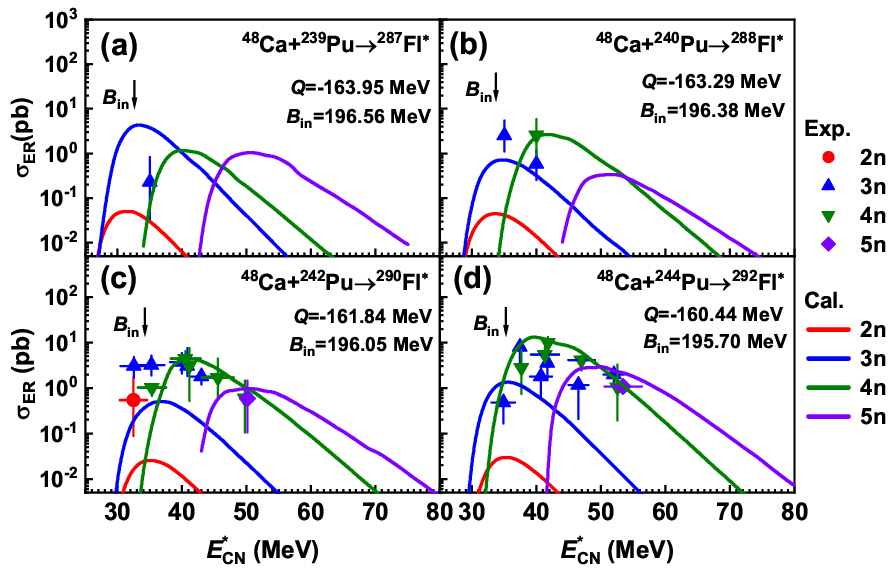}
	\caption{\label{fig7} 
		Same as Fig. \ref{fig6}, but for hot-fusion reactions with target nuclei $^{239,240,242,244}$Pu, juxtaposed with experimental data indicated by error bars \cite{PhysRevLett.83.3154,PhysRevC.62.041604,PhysRevC.70.064609,PhysRevC.92.034609,PhysRevC.97.014320}.
		The calculated ERCS and experimental data for the 5n-channel are represented by purple solid lines and purple solid diamonds, respectively.
	}
\end{figure}

\begin{figure}[htbp]
	\includegraphics[width=1.\linewidth]{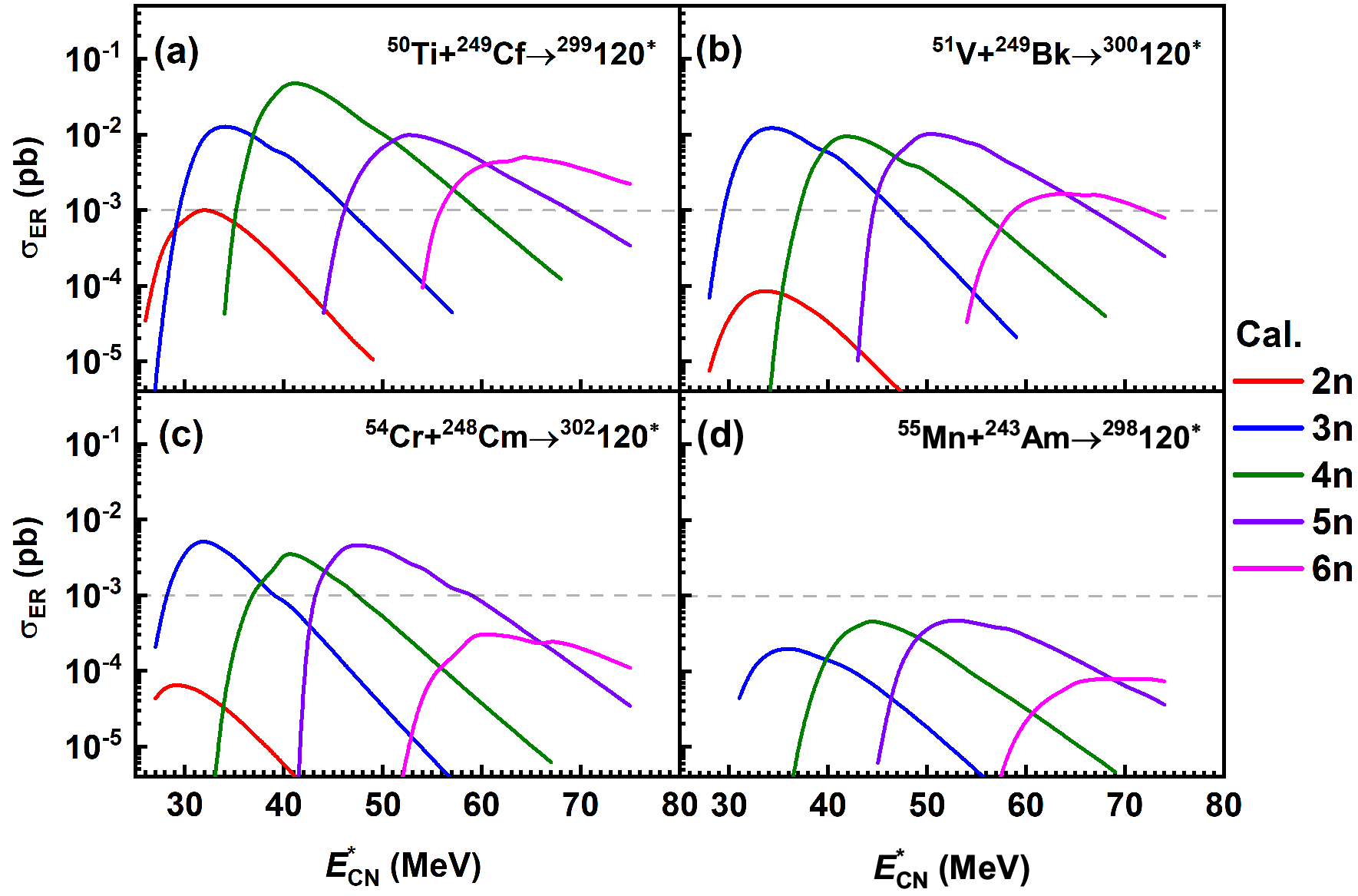}
	\caption{\label{fig8} 
		Panels (a) to (d) show the predicted excitation functions of ERCS for the reactions $^{50}$Ti+$^{249}$Cf, $^{51}$V+$^{249}$Bk, $^{54}$Cr+$^{248}$Cm, and $^{55}$Mn+$^{243}$Am, respectively.
		The calculated ERCS for the 2n-, 3n-, 4n-, 5n- and 6n-channels are represented by red, blue, olive, purple, magenta solid lines, respectively.
	}
\end{figure}

Based on the DNS model and incorporating the microscopic input physical quantities including the single neutron separation energies $S_{\mathrm{n}}$, fission barriers $B_{\mathrm{f}}(E^*)$, shell correction energies $E_{\mathrm{sh}}$, as well as the asymptotic level density parameters $\tilde{a}$, we perform calculations for several systems. These include the cold fusion reactions of $^{48}$Ca + $^{204,206-208}$Pb, presented in Fig. \ref{fig6}, and the hot fusion reactions of $^{48}$Ca + $^{239,240,242,244}$Pu, shown in Fig. \ref{fig7}. Additionally, calculations are carried out for reactions aimed at synthesizing the unknown superheavy element 120—specifically, $^{50}$Ti + $^{249}$Cf, $^{51}$V + $^{249}$Bk, $^{54}$Cr + $^{248}$Cm, and $^{55}$Mn + $^{243}$Am—as shown in Fig. \ref{fig8}. It is noteworthy that through the comparison of the three methods for extracting the level density parameter, the third method ($a_{3}=(1/2)(\partial S/\partial T)$) is found to be more suitable for the DNS model.

Figure \ref{fig6} shows the excitation function of the production cross section for the synthesis of the nucleus No in cold fusion reactions, comparing theoretical calculations with experimental data \cite{GAGGELER1989561,belozerov2003spontaneous,PhysRevC.69.054607}.
Fig. \ref{fig6} contains four panels: (a) $^{48}$Ca + $^{204}$Pb, (b) $^{48}$Ca + $^{206}$Pb, (c) $^{48}$Ca + $^{207}$Pb, and (d) $^{48}$Ca + $^{208}$Pb.   
The characteristic of cold fusion reactions is that the peak of the excitation function is near the Coulomb barrier, usually corresponding to the 1n- or 2n-channel. This is mainly because the projectile or target nuclei have strong shell effects, resulting in a smaller reaction $Q$ value.
Our calculations show good agreement with the experimental data, particularly in the region around the Coulomb barrier. However, at higher excitation energies, the calculated values exceed the experimental results. The primary reason for this discrepancy is that the fission barriers obtained from the microscopic model fade out too slowly as the excitation energy increases. Consequently, the fission barriers become higher than those that fade out rapidly. As clearly shown in Fig. \ref{fig6}, the peak value of the excitation function of the ERCS exhibits a strong dependence on the target isotopes. This is because neutron-deficient target isotopes tend to form more neutron-deficient compound nuclei, which have larger neutron-separation energies and lower fission barriers, resulting in reduced survival probability.

Figure \ref{fig7} shows the excitation function of ERCS for the synthesis of the nucleus Fl in hot fusion reactions, comparing theoretical calculations with experimental data \cite{PhysRevLett.83.3154,PhysRevC.62.041604,PhysRevC.70.064609,PhysRevC.92.034609,PhysRevC.97.014320}.
Fig. \ref{fig7} contains four panels: (a) $^{48}$Ca + $^{239}$Pu, (b) $^{48}$Ca + $^{240}$Pu, (c) $^{48}$Ca + $^{242}$Pu, and (d) $^{48}$Ca + $^{244}$Pu. 
The characteristic of hot fusion reactions is that the peak of the excitation function is near the Coulomb barrier, slightly above it, and usually corresponds to the 3n- or 4n-channel. Compared with cold fusion, the shell effect in the $Q$ value of hot fusion reactions is weaker, hence its $Q$ value is larger.
Our calculations show good agreement with the experimental data, particularly in panels (c) and (d). Due to the limited experimental data for the reactions \(^{48}\text{Ca} + ^{239}\text{Pu}\) and \(^{48}\text{Ca} + ^{240}\text{Pu}\) shown in panels (a) and (b), although it is not possible to determine their excitation functions, a systematic comparison with the calculated results can still be made. 
Compared to cold fusion reactions in Fig. \ref{fig6}, the peak values of excitation functions of ERCS for hot fusion reactions of \(^{48}\text{Ca} + \text{Pu}\) have weak dependence on the target isotopes, because high excitation energy reduces the difference in fission barriers between isotopes of the compound nucleus.

Considering the availability of projectile-target combinations, four reaction systems $^{50}$Ti+$^{249}$Cf, $^{51}$V+$^{249}$Bk, $^{54}$Cr+$^{248}$Cm, and $^{55}$Mn+$^{243}$Am are selected in Fig. \ref{fig8} for the synthesis of superheavy element 120.
Fig. \ref{fig8} shows the calculated ERCS as functions of excitation energies via hot fusion reactions, based on the DNS model, incorporating the microscopic input physical quantities.
Since the peak values of the excitation functions for the ERCS correspond to the 3$\sim$5n-channels, these four reactions are classified as hot fusion reactions.
As clearly shown in Fig. \ref{fig8}, the peak values of the excitation functions gradually decrease with increasing excitation energy, owing to the slower decrease of the fission barrier. From panel (a) to (d), the peak values corresponding to each reaction system gradually decrease. Compared to the reaction in panel (a), the peak values in panel (d) are reduced by approximately two orders of magnitude. This suppression can be attributed to the high symmetry between the projectile and target nuclei. In general, when forming similar superheavy element, the use of a heavier projectile reduces the fusion probability, as it requires the transfer of a larger number of nucleons to complete the fusion process.
As shown in Fig. \ref{fig8}, the maximum ERCS and corresponding excitation energies for the 3n-, 4n-, and 5n-channels in each reaction systems are identified as shown in TABLE \ref{tab2}.
\begin{table}[htbp]
	\renewcommand{\arraystretch}{1.19}
	\centering
	\setlength{\tabcolsep}{6pt}
	\caption{\label{tab2} The maximum ERCS and corresponding excitation energies for the 3n-, 4n-, and 5n-channels in the reaction systems $^{50}$Ti+$^{249}$Cf, $^{51}$V+$^{249}$Bk, $^{54}$Cr+$^{248}$Cm, and $^{55}$Mn+$^{243}$Am.}
	\begin{tabular}{rcc}
		\toprule
		Reactions\hspace{3em} & $\sigma_{\rm ER}$(fb) & $E^*_{\rm CN}$(MeV)  \\
		\hline
		$^{50}$Ti($^{249}$Cf,3n)$^{296}$120 & 12.87 & 34   \\
		$^{50}$Ti($^{249}$Cf,4n)$^{295}$120 & 48.20 & 41    \\
		$^{50}$Ti($^{249}$Cf,5n)$^{294}$120 & 9.90 & 52    \\
		$^{51}$V($^{249}$Bk,3n)$^{297}$120  & 12.33 & 34   \\
		$^{51}$V($^{249}$Bk,4n)$^{296}$120  & 9.51 & 42   \\
		$^{51}$V($^{249}$Bk,5n)$^{295}$120  & 10.18 & 50   \\
		$^{54}$Cr($^{248}$Cm,3n)$^{299}$120 & 5.25 & 32     \\
		$^{54}$Cr($^{248}$Cm,4n)$^{298}$120 & 3.55 & 40     \\
		$^{54}$Cr($^{248}$Cm,5n)$^{297}$120 & 4.60 & 47     \\
		$^{55}$Mn($^{243}$Am,3n)$^{295}$120 & 0.20 & 36     \\
		$^{55}$Mn($^{243}$Am,4n)$^{294}$120 & 0.46 & 44     \\
		$^{55}$Mn($^{243}$Am,5n)$^{293}$120 & 0.47 & 53     \\
		
		\bottomrule
	\end{tabular}
\end{table}
Among these, the reaction $^{50}$Ti + $^{249}$Cf exhibits the most promising projectile-target combination for the synthesis of element 120.
It is found that the optimum excitation energies for these four reaction systems show large differences, which are primarily attributed to the fission barrier and the neutron separation energy. First, the temperature-dependent fission barrier decreases slowly at excitation energies above 30 MeV, resulting in the slight difference between the peak values for the (3-5)n channels. Additionally, the subtle variations in the neutron separation energies can lead to substantial differences in survival probabilities.

\section{Summary}\label{sec4}

Our previous studies demonstrated that the predictive accuracy of the DNS model for superheavy element synthesis cross-sections is highly sensitive to the precision of its input parameters, including nuclear masses and related quantities. In the present work, to achieve self-consistency, we generate essential input quantities—such as neutron separation energies, fission barriers, shell correction energies, asymptotic level density parameters, and shell damping factors—within a unified framework based on finite-temperature CDFT using the PC-PK1 energy density functional, with pairing correlations treated via the BCS approach.
Owing to the considerable computational cost associated with this microscopic model, calculations are restricted to selected isotopic chains rather than the entire region of SHN. Accordingly, we focus on three representative elements: No, Fl, and unknown superheavy element 120, whose isotopic physical quantities are listed in TABLE \ref{tab1}. Among these, No and Fl have experimentally measured data, allowing for validation and constraints on the DNS model. For the reactions $^{48}$Ca + $^{204,206–208}$Pb and $^{48}$Ca + $^{239,240,242,244}$Pu, our calculations show good agreement with experimental excitation functions. However, the theoretical excitation functions of the ERCS decrease more gradually than experimental observations, which we attribute to the slower decrease of the fission barrier.
Furthermore, we predict the synthesis cross sections of element 120 via the following reactions: $^{50}$Ti + $^{249}$Cf, $^{51}$V + $^{249}$Bk, $^{54}$Cr + $^{248}$Cm, and $^{55}$Mn + $^{243}$Am. It is found that the optimal channels are identified as: $^{50}$Ti($^{249}$Cf,4n)$^{295}$120 with $\sigma_{\rm ER} = 48.20$ fb at $E^*_{\rm CN} = 41$ MeV; $^{51}$V($^{249}$Bk,3n)$^{297}$120 with $\sigma_{\rm ER} = 12.33$ fb at $E^*_{\rm CN} = 34$ MeV; $^{54}$Cr($^{248}$Cm,3n)$^{299}$120 with $\sigma_{\rm ER} = 5.25$ fb at $E^*_{\rm CN} = 32$ MeV; and $^{55}$Mn($^{243}$Am,5n)$^{293}$120 with $\sigma_{\rm ER} = 0.47$ fb at $E^*_{\rm CN} = 53$ MeV. It is expected that the results of these calculations will provide theoretical guidance for future experiments aimed at synthesizing element 120.

\section*{Acknowledgments}
Helpful discussions with Wei Gao are gratefully acknowledged. This work is supported by National Science Foundation of China (NSFC) (Grant No. 12105241) and the Natural Science Foundation of Henan Province (Grant No. 262300421888).
\end{CJK}
\vspace{-1mm}
\centerline{\rule{80mm}{0.1pt}}
\vspace{2mm}



\clearpage

\end{document}